\documentclass[preprint,12pt]{elsarticle}

\usepackage{epsfig,amsmath,amssymb,natbib,color}

\usepackage{xcolor}

\journal{}

\begin{document}

\begin{frontmatter}



\title{Influence of a vertical-wall leading edge on bouncing and escape bubble rising regimes} 

\author[label1,label2]{A. Rubio-Gonz\'alez\corref{cor1}}
\author[label1,label2]{E. J. Vega}
\author[label3,label4]{R. Bolaños-Jim\'enez}

\address[label1]{Depto. de Ingenier\'{\i}a Mec\'anica, Energ\'etica y de los Materiales, Universidad de Extremadura, E-06006 Badajoz, Spain}
\address[label2]{Instituto de Computaci\'on Cient\'{\i}fica Avanzada (ICCAEx), Universidad de Extremadura, E-06006 Badajoz, Spain}
\address[label3]{Depto. de Ingenier\'{\i}a Mec\'anica y Minera, \'Area de Mec\'anica de Fluidos, Universidad de Ja\'en, Campus de las Lagunillas, 23071 Ja\'en, Spain}
\address[label4]{Andalusian Institute for Earth System Research, Campus de las Lagunillas, 23071 Ja\'en, Spain}
\cortext[cor1]{Corresponding author: A. Rubio-Gonz\'alez, \texttt{arubiorg@unex.es}}




\begin{abstract}

This work investigates deformable gas bubbles rising near a vertical wall in ultrapure water, focusing on how the position of the wall leading edge affects their near-wall dynamics. Two configurations are considered: (i) a wall extending from 11--25 bubble diameters below the bubble injection point, so that the bubble rises under the continuous influence of the boundary; and (ii) a wall whose leading edge is located 144 mm above the injector, corresponding to approximately 80--180 bubble diameters, allowing the bubble to reach its terminal velocity before entering the wall-bounded region. The results show that the wall leading-edge position influences both the transition from periodic bouncing (PB) to bouncing--tumbling--escaping (BTE) dynamics and the rebound frequency within the PB regime. When the wall leading edge is placed downstream, the onset of BTE occurs at smaller bubble sizes, corresponding simultaneously to lower Bond ($Bo$), Galilei ($Ga$), and Reynolds ($Re$) numbers. The Strouhal number ($St$) follows a similar decreasing trend in both configurations at low Bond numbers, but the two behaviours diverge for ($Bo \gtrsim 0.15$). For the PB regime, when the wall extends from the injector, $St$ approaches an approximately constant value of $St\simeq0.014$, whereas substantially lower values are measured for the downstream-wall configuration. The larger rebound amplitudes and longer return stages observed in the latter configuration account for lower frequencies. These findings indicate that the bubble dynamics depend not only on the conventional control parameters and wall separation, but also on the wall geometry and the associated pre-interaction bubble hydrodynamic history.

\end{abstract}


\end{frontmatter}



\section{Introduction}
\label{sec1}

The rising of a single bubble in an unbounded, quiescent fluid is typically captured by two dimensionless numbers: the Bond number ($Bo = \rho g D^{*2} / \sigma$), and the Galilei number ($Ga = \sqrt{g D^{*3}}  \rho / \mu$), with $D^{*}$ the equivalent bubble diameter (the asterisk denotes dimensional magnitudes), $\rho$ and $\mu$ the liquid density and viscosity, respectively, $\sigma$ the surface tension, and $g$ the gravity acceleration. The combination of these two parameters determines the resulting Reynolds number ($Re=\rho v_t^{*} D^{*}/\mu$, with $v_t^{*}$ the bubble terminal velocity), and the bubble major-to-minor diameter ratio ($\chi$), called aspect ratio. At low $Bo$, the bubble remains nearly spherical, and exhibits a steady, rectilinear trajectory~\citep{CGW78,BM95}. As $Bo$ and $Ga$ increase, the bubble becomes more deformed and the wake undergoes a transition from steady to unsteady, leading to path instabilities such as zigzagging or spiraling~\citep{MM02,YPT03,ZM08,CMMT16}. These instabilities are associated with periodic shedding of vortical structures, which vary in symmetry and intensity with bubble shape and velocity, and alter the terminal velocity and lead to abrupt transitions in the bubble’s dynamical regime~\citep{TSG15,CMMT16}.

Most practical situations involve bubbles rising near solid boundaries, such as walls or immersed objects, which significantly affect their motion. This proximity breaks the axial symmetry of the wake, altering pressure and viscous forces and resulting in lateral forces that may either attract or repel the bubble from the wall~\citep{M03c,ST10,VBW02,ZDCY20,ECMBBJ25}. The interplay between the bubble and a nearby vertical wall is controlled by two opposing forces depending on the value of the control parameters~\citep{SZM24,Shi24}. The first is an attraction due to the Bernoulli effect, where faster liquid flow in the space between the bubble and the wall pulls them together due to the existence of a pressure minimum in the gap. The second is a repulsion caused by the bubble's wake interacting with the wall, due to vorticity asymmetries across the bubble caused by the presence of the wall for moderate-to-high $Re$, while the physical mechanism responsible for the migration at low $Re$ is fundamentally due to the asymmetric viscous diffusion generated on the surface~\citep{TTMM02,TM03}. For slightly deformed bubbles and low inertial regimes ($Re \lesssim 35$), the repulsive effect prevails, while for higher inertial regimes, the rising path is a consequence of the competition between both the attractive Bernoulli effect and the repulsive force due to the vortical mechanism. As $Re$ increases, the attractive irrotational component becomes significant, often leading to a bouncing behaviour where the bubble alternatively approaches and recoils from the wall. This balance between attractive and repulsive forces has also been explored numerically by \citet{ST15}, who confirmed the presence of both effects and emphasised the importance of accounting for bubble deformation and wall-induced wake asymmetries.

Very recently, \citet{SZM24,SZM25} conducted numerical studies focused on the effect of wall proximity and confinement on bubbles undergoing wake-induced instabilities for moderate-to-high inertial regimes.
Their results confirm that the presence of a wall promotes vortex shedding and the transition to unsteady regimes, and highlight that the magnitude and spatial distribution of the repulsive force strongly depend on the confinement geometry and the relative bubble size. Moreover, they identified several regimes of bubble motion in the ($Ga, Bo$) plane for a gas injector-to-wall separation of one diameter, $L = y_w^{*}/D^{*}=1$. They showed that when $Ga$ and $Bo$ remain below critical thresholds, bubbles may undergo periodic bouncing near the wall or escape the near-wall region after rebounding. A periodic bouncing (PB) regime results from the competition between the Bernoulli attraction generated in the narrow gap and an unsteady repulsive force associated with the generation and shedding of streamwise vortices along the wall. The bubble is able to escape when the collision generates a sufficiently intense rotational flow around the bubble, producing a Magnus-like lift force that overcomes the wall attraction.

\citet{GRBV26} extend and complement these results experimentally at moderate-to-high $Re$ for a varying value of $L$, specifically, they characterized the effect of $L$ on the regime map $(Ga,Bo)$ when the leading edge of the wall is located downstream from the gas injector. The experiments revealed the existence of four distinct behaviours, \textit{i.e.}, rectilinear path (RP), periodic bouncing or collisions (PB), migration away (MA), and collision and migration away (C+MA), in agreement with recent numerical simulations~\cite{SZM24,SZM25} mentioned above. 

It should be noted that C+MA regime is denoted as BTE (Bouncing–Tumbling–Escaping) regime in numerical studies, and it will be referred to by this designation hereafter. Nevertheless, although an overall qualitative agreement in the classification of regimes between experiments and simulations, discrepancies were observed regarding the onset of escape-type dynamics after collision. In particular, the onset of the BTE regime was established at lower values of $Ga$ and $Bo$ than those in the numerical simulations. The observed differences were attributed to the experimental configuration, in which the wall was positioned downstream of the injector, allowing the bubble to reach its terminal velocity and develop an approximately axisymmetric wake before interacting with the solid boundary. In contrast, in the numerical studies, the wall was present from the beginning of the rise, so the wake evolved under wall influence from the onset. 

Despite these advances, the role of the wall leading edge has not yet been isolated. Therefore, the aim of the present work is to investigate the role of the vertical distance between the injector and the wall edge in determining the bubble rising regime. In particular, we focus on its effect on the PB-BTE transition and on the bouncing frequency in the PB regime. 

The paper is organised as follows. Section \ref{sec2} presents the experimental method followed to obtain the results concerning the effect of the wall-edge vertical position on the regimes, and the bouncing frequency and amplitude are shown in Section \ref{sec3}. Finally, Section \ref{sec4} is devoted to the main conclusions of the work. 

\section{Experimental method} 
\label{sec2}

The experimental setup is sketched in Figure \ref{setup}. We used a parallelepipedal glass tank/column (A in Fig.~\ref{setup}a) to ensure optical access and maintain cleanliness, measuring 10 cm in both length and width and 100 cm in height. As liquid, we used ultrapure water in all our experiments, provided by a water purification machine ({\sc Direct-Q®3}). The density and viscosity were $\rho=998$ kg/m$^3$ and $\mu=1$ mPa$\cdot$s at $T=293\pm1$ K, with a conductivity of around 3.5 $\mu$S/cm. The gas used in the experiments was nitrogen (=99.998\%, $\rho_g=1.25$ kg/m$^3$, $\mu_g=17.6$ $\mu$Pa$\cdot$s, {\sc Carburos Metálicos SL}). The surface tension of the gas-liquid interface was $\sigma=72$ mN/m, measured using the Theoretical Interface Fitting Analysis ({\sc TIFA})~\citep{FMC07}. To ensure quasi-static bubble formation, bubbles were generated by injecting nitrogen at a low enough constant flow rate through a needle placed at the centre of the tank base (B in Fig.~\ref{setup}a). The needle tips were fabricated using Nanoscribe Photonic Professional GT2 with the dip-in laser lithography (DiLL) configuration ~\citep{RVCMLH24}, with inner diameters ranging from 10 to 90 $\mu$m to control the bubble size.

\textbf{\begin{figure}[tbp]
\begin{center}
\resizebox{1\textwidth}{!}{\includegraphics{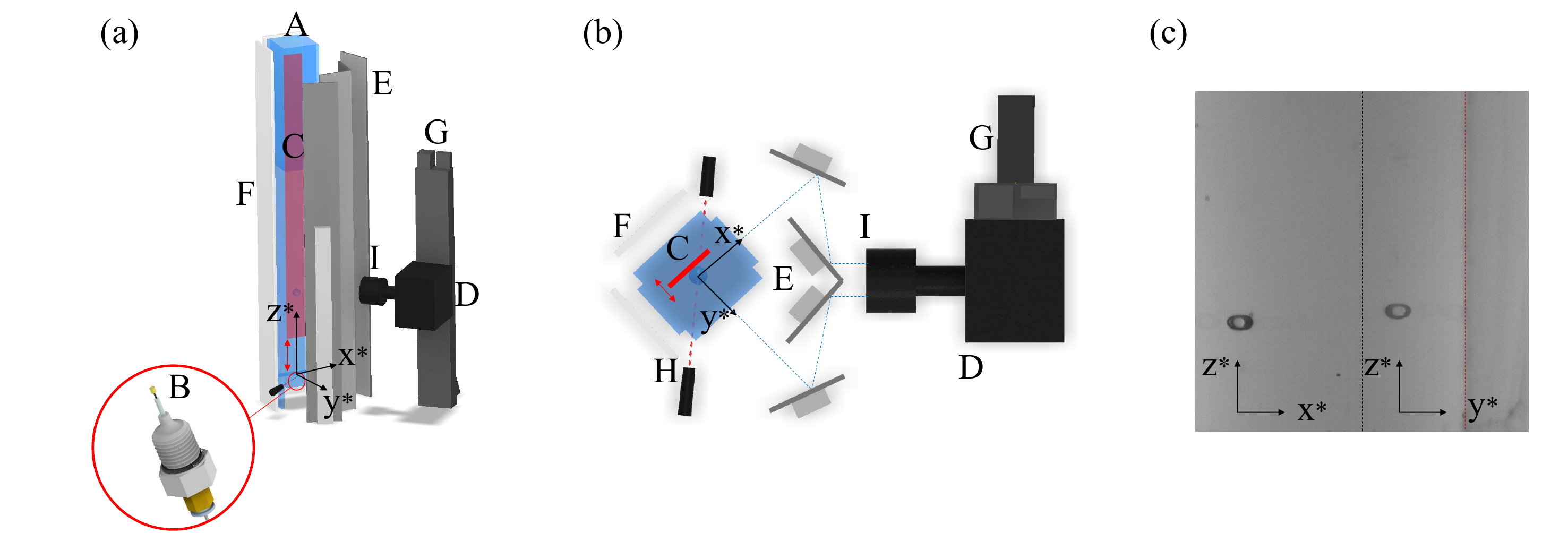}}   
\end{center}
\caption{Experimental setup: three-dimensional scheme (a), top view (b), real image of the experiment (c). Components: tank (A), needle (B), mobile glass wall (C), high-speed camera (D), sets of mirrors (E), LED panel lights (F), vertical motorised rail (G), photo-diode sensor (H), and optical lenses (I). The image on the right side shows the two perpendicular views of the bubbles at the moment when the wall appears in the two views.}
\label{setup}
\end{figure}}

A rigid and hydrophilic glass wall ($80\times1000\times6$ mm, with a typical root-mean-square roughness of a few nanometres) was placed vertically inside the tank (C in Figs.~\ref{setup}a,b). The wall was ensured to be parallel to the $x^*z^*$ plane using a goniometer with micrometric precision. The vertical and horizontal distances from the wall to the centre of the injector, that is, to the bubble centroid just after its generation, $z_w^{*}$ and $y_w^{*}$, respectively, were controlled with a biaxial translation stage with micrometric precision. Two configurations were considered in order to assess the influence of the wall leading-edge position on bubble-wall interactions. In the first configuration, the wall extends from the injection point (called $z_w^* = 0$ mm herein, although the wall edge was positioned more than 11 equivalent diameters below the gas injector tip, $11\lesssim z_w=z_w^*/D \lesssim 25$, for all bubbles investigated, to avoid its influence during bubble growth and detachment), corresponding to the arrangement commonly adopted in previous numerical studies. In the second configuration, the wall is located downstream of the injector, with the leading edge positioned at $z_w^*=144$ mm. This position corresponds to more than 80 equivalent diameters ($80\lesssim z_w=z_w^*/D \lesssim 180$). This vertical distance was selected based on previous measurements~\citep{GRBV26}, which showed that the bubble velocity had already converged to its terminal value long before reaching the wall position adopted here. Although this downstream-wall configuration was previously examined by~\citet{GRBV26}, in the present study, numerous additional experiments were carried out to enable a consistent comparison with the $z_w^* = 0$ mm case, and to systematically analyse the influence of the wall leading edge on the bubble's rising regime and its bouncing dynamics. In all experiments, the horizontal wall position was adjusted so that the dimensionless gas injector-wall horizontal separation was $L=1$. 

To perform the measurements, a virtual binocular stereo vision system~\citep{LWZGZXLL22, RVCMLH24} was used to capture two perpendicular views of the rising bubble. The optical and acquisition system was identical to that described in ~\citep{GRBV26}, and only a brief overview is provided here. A  high-speed CMOS camera ({\sc Photron, Fastcam Mini UX50}) simultaneously captured two perpendicular views of the bubble (D in Figs.~\ref{setup}a,b), using a mirror arrangement illustrated in Fig. \ref{setup} (E in Figs.~\ref{setup}a,b). The experiment illumination was provided by two LED panels with a diffuser to ensure uniform backlighting (F in Figs.~\ref{setup}a,b). The camera was mounted on a vertical motorised rail (G in Figs.~\ref{setup}a,b) moving at constant speed to track the bubble ascent, triggered at pinch-off by a laser-photodiode system (H in Fig.~\ref{setup}b). Recordings were performed at 500 fps with a 625 $\mu$s exposure time and a resolution of 1024×1280 pixels. The optical configuration includes a {\sc Nikon} zoom lens, a 2$\times$ zoom, and 20 mm extension rings (I in Fig.~\ref{setup}b), achieving a magnification of 21.62 $\mu$m/pixel. All components were mounted on an optical table equipped with a pneumatic anti-vibration isolation system ({\sc Thorlabs}) to eliminate external vibrations from the building. All experiments were carried out at room temperature (20 $^{\circ}$C). The experimental procedure is similar to that described in ~\citet{GRBV26}, to which the reader is referred for further details. Briefly, before each experimental run, the appropriate injector tip was installed, and the tank was thoroughly rinsed with ultrapure water prior to filling with the working liquid. Each experiment was repeated at least 5 times.

The bubble rise dynamics depend on the equivalent diameter $D^{*}$, the liquid properties $(\rho,\mu,\sigma)$, gravity $g$, the initial wall horizontal distance $y_w^{*}$, and the vertical position of the wall leading edge, $z_w^*$. Bubble size is characterised by the equivalent diameter $D^{*}=(6V^{}/\pi)^{1/3}$. In the present experiments, $0.58 \le D^{*} \le 1.76$ mm, below the critical diameter for path instability, ensuring a stable trajectory in the absence of a wall. Using $D^{*}$ and $(gD^{*})^{1/2}$ as characteristic length and velocity scales, respectively, the problem is governed by three dimensionless parameters: the Galilei number $Ga=\rho (g D^{*3})^{1/2}/\mu$, the Bond number $Bo=\rho g D^{*2}/\sigma$, and the dimensionless injector-to-wall horizontal distance $L=y_w^{*}/D^{*}$, together with the dimensionless height of the wall leading edge, $z_w$. The latter, as reported before, was set in two positions, namely $z_w=0$ and $z_w>80$. The Morton number $Mo=g\mu^{4}/(\rho\sigma)^{3}$ remains constant ($Mo=2.64\times10^{-11}$), linking $Bo$ and $Ga$ through $Bo=Mo^{1/3}Ga^{4/3}$. The explored ranges are $0.045\lesssim Bo\lesssim0.42$, $43\lesssim Ga\lesssim231$, and $50\lesssim Re_D\lesssim624$. 

Images were processed at the pixel level using an in-house \textsc{Matlab} code to obtain the bubble centroid, shape, and velocity (see also the supplementary material of ~\citet{RVCMLH24}). The deformation was quantified by the aspect ratio $\chi(t)$ (major-to-minor diameter), computed from both camera views. Each case was repeated five times; reported values correspond to ensemble averages, with uncertainty given by the standard deviation. The deformation range was $1.02\lesssim\chi\lesssim1.86$, and trajectories were highly reproducible across realizations.

\section{Results and discussion} 
\label{sec3}
In this Section, we present the experimental results. The influence of the wall leading edge is examined, first on the bubble rising regime in Subsection~\ref{subsec:regime}, and subsequently on the bouncing frequency and amplitude within the PB regime in Subsection~\ref{subsec:PB}. 

\subsection{Effect of the wall leading edge on the rising regime} \label{subsec:regime}

\begin{figure}[tbp]
\begin{center} 
\resizebox{0.95\textwidth}{!}{\includegraphics{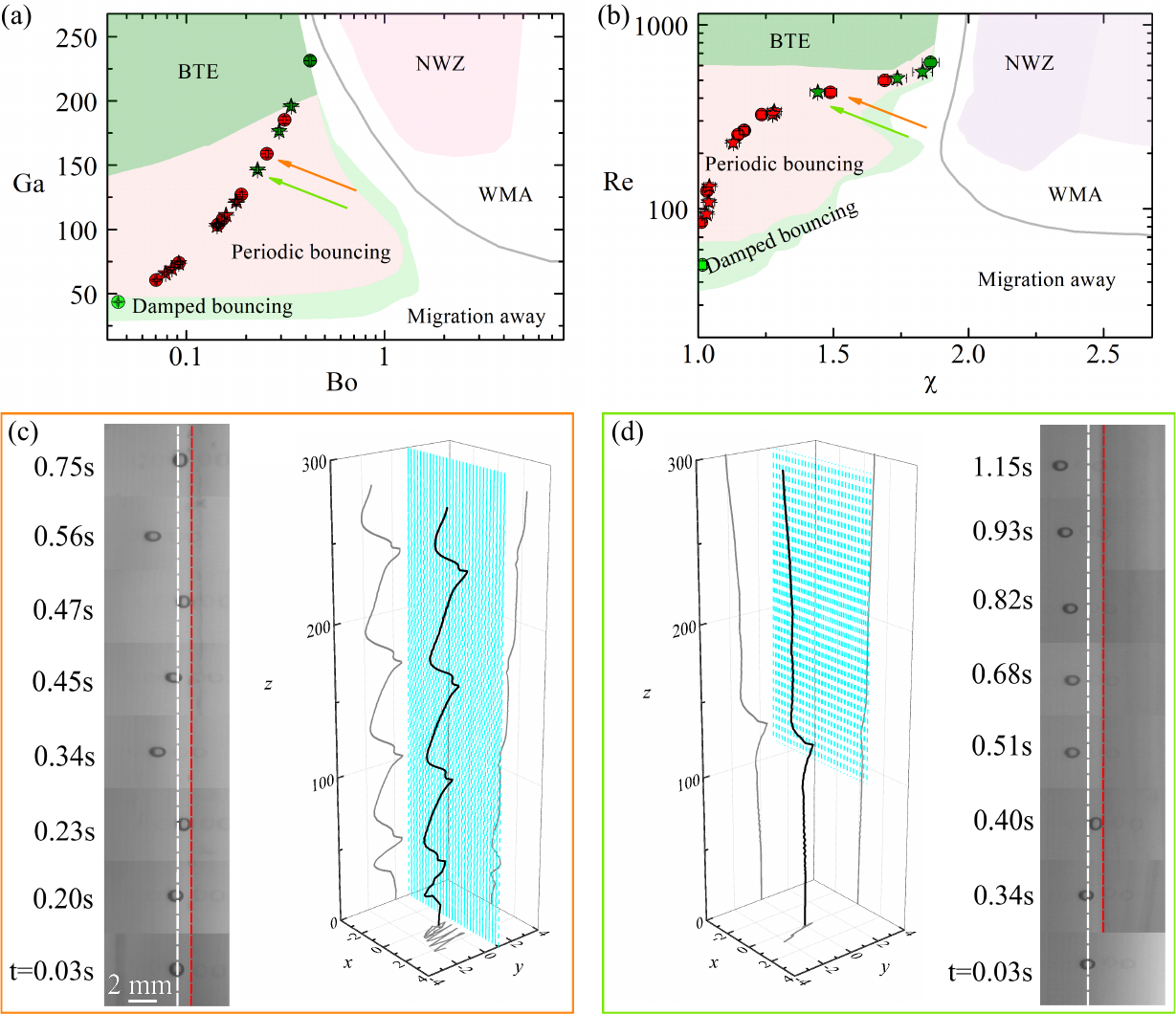}}
\end{center} 
\caption{Experimental cases investigated in the present study ($L=1$) for a wall present from the onset ($z_w^*=0$ mm, circles) and located downstream ($z_w^*=144$ mm, stars), superimposed on the (a) $Bo-Ga$ regime map and (b) $\chi-Re$ regime map reported by \citet{SZM25} for $z_w^*=0$. Red, light green and dark green symbols correspond to experiments for PB (Periodic bouncing), DB (damped bouncing), and BTE (bouncing--tumbling--escaping) behaviours, respectively. The coloured background and regime boundaries correspond to the numerical results and are included for reference. The solid line is the neutral curve corresponding to the onset of path instability in an unbounded fluid~\citep{BSFM24}. The white region straddling the neutral curve represents the whole set of conditions under which bubbles migrate away from the wall, either in the presence or in the absence of path instability, and NWZ region means near-wall zigzagging. The cases analyzed in panels (c) and (d) are highlighted by the orange and green arrows, respectively. (c,d) Representative time sequences and corresponding dimensionless bubble trajectories for the $z_w^* = 0$ mm and $z_w^* = 144$ mm cases, respectively. Both cases correspond to the same flow conditions, $Ga\approx152$, $Bo\approx0.24$, $Re\approx420$ and $\chi\approx1.5$. In the image sequences, the white dashed line marks the $y=0$ reference and spans the streamwise $z$ direction, while the red dashed line indicates the leading edge of the wall.}
\label{fig:map_cases}
\end{figure}

Figures \ref{fig:map_cases}(a)-(b) present the experimental regimes observed in the present study for $z_w^*=0$ (circles) and $z_w^*=144$ mm (stars), superimposed on the regime map reported by \citet{SZM25} for bubbles rising near a vertical wall located at $L=1$ and extending from the injection point ($z_w^*=0$). Since all experiments were conducted in ultrapure water ($Mo=2.64\times10^{-11}$), the accessible parameter space is restricted to a limited region of the complete map. Within this region, the experimental cases span the transitions from damped bouncing dynamics to periodic bouncing or collisions (PB), and from PB to BTE regime. For the smallest Bond numbers investigated, the bubble remains trapped in the vicinity of the wall and exhibits bounded oscillatory motion. As the Bond and Galilei numbers increase, the interaction becomes progressively more energetic until, above a critical threshold, the bubble escapes from the near-wall region following collision with the wall, corresponding to the onset of the BTE regime. Note that the resulting experimental regimes for $z_w^*=0$ (circles) are qualitatively consistent with the numerical results reported by \citet{SZM24,SZM25} for the same wall configuration. In particular, the transition from DB to PB is observed experimentally at $Bo_{c,exp}=0.058$ and $Re_{c,exp}=67$, compared with $Bo_{c,num}=0.05$ and $Re_{c,num}=66$ in the numerical simulations. Likewise, the onset of the BTE regime occurs at $Bo_{c,exp}=0.34$ and $Re_{c,exp}=561$ in the experiments and at $Bo_{c,num}=0.35$ and $Re_{c,num}=570$ in the simulations. In those simulations, the observed transitions were interpreted as the result of the competition between the attractive Bernoulli mechanism associated with the accelerated flow in the gap and wall-induced vortical effects that promote lateral migration away from the wall. Although the present experiments do not provide direct measurements of the liquid flow field and therefore cannot verify these mechanisms, the overall agreement between both studies suggests that similar physical processes are likely to govern the observed transitions. Therefore, the experimental observations confirm the occurrence of the periodic-bouncing regime within the region predicted numerically. 

To investigate how the position of the wall leading edge modifies the transition between the bouncing and escaping regimes, the regime map in the reference configuration ($z_w^*=0$) is compared with that in a second configuration in which the wall begins downstream of the injector ($z_w^*=144$ mm). In the following, the previous $z_w^*=0$ configuration is used as a baseline against which the downstream-wall configuration is compared in order to assess how the presence of a finite wall leading edge modifies the transition boundaries between the bouncing and the escaping regimes. The most notable change concerns the transition between the PB and BTE regimes. When the wall is present from the injection point ($z_w^*=0$), periodic collisions persist over a substantially larger portion of the parameter $Ga-Bo$ and $Re-\chi$ spaces. In contrast, when the wall is located downstream, the transition is shifted towards smaller bubble sizes along the fixed-$Mo$ experimental path, corresponding simultaneously to lower $Bo$, $Ga$, $Re$ and $\chi$. Particularly, the critical Bond and Reynolds numbers shift from $Bo_{c,0}=0.34$ and $Re_{c,0}=561$ for the $z_w^*=0$ configuration to $Bo_{c,144}=0.20$ and $Re_{c,144}=384$ for the $z_w^*=144$ mm configuration. Consequently, bubbles that exhibit sustained periodic collisions in the reference configuration may escape from the near-wall region after a single collision when the interaction takes place only after a free-rise stage. In fact, the influence of the wall leading-edge position becomes evident in Figures~\ref{fig:map_cases}(c),(d), showing the path of two bubbles with similar $Bo$ and $Re$, but exhibiting PC regime when $z_w^*=0$ (Fig.~\ref{fig:map_cases}c) and BTE regime when $z_w^*=144$ mm (Fig.~\ref{fig:map_cases}d). 

According to the DNS of \citet{SZM24,SZM25}, the PB regime results from the interplay between the attractive gap-induced pressure force and an unsteady repulsive contribution associated with the generation and shedding of streamwise vortices.

In contrast, the BTE regime develops when the rotational flow generated during collision becomes sufficiently intense to produce a Magnus-like lift force capable of driving the bubble away from the wall.

Although the present experiments do not provide direct measurements of the liquid velocity field, the observed shift of the PB–BTE boundary suggests that the wall leading-edge position modifies the conditions under which these mechanisms develop. When the wall extends from the bubble injection point, the bubble rises under the continuous influence of the wall, so the near-wall shear and wake asymmetry develop gradually during the entire ascent. By contrast, in the downstream-wall configuration, the bubble first rises freely and reaches its terminal state before suddenly entering the wall region. The subsequent interaction therefore starts from a different hydrodynamic condition: the wake develops without the constraint imposed by the wall and must reorganise abruptly once the bubble encounters the solid boundary. This abrupt reorganisation is consistent with the earlier onset of the BTE regime observed experimentally and may result from an enhancement of the rotational flow generated during the first collision, although direct flow-field measurements would be required to confirm this mechanism. 

These observations indicate that the wall leading-edge position modifies the hydrodynamic state of the bubble–wake system before the first collision. As a consequence, the conditions under which the collision-induced rotational flow develops are also modified, shifting the PB–BTE transition towards lower values of $Bo$, $Ga$, $Re$ and $\chi$. In this sense, the wall leading edge should be regarded not merely as a geometric feature, but as a parameter controlling the wake development before the impact. The good agreement obtained for the $z_w^*=0$ configuration further supports the consistency between the present experiments and the DNS when equivalent boundary conditions are considered.

\begin{figure}[htbp]
\begin{center}
\resizebox{0.95\textwidth}{!}{\includegraphics{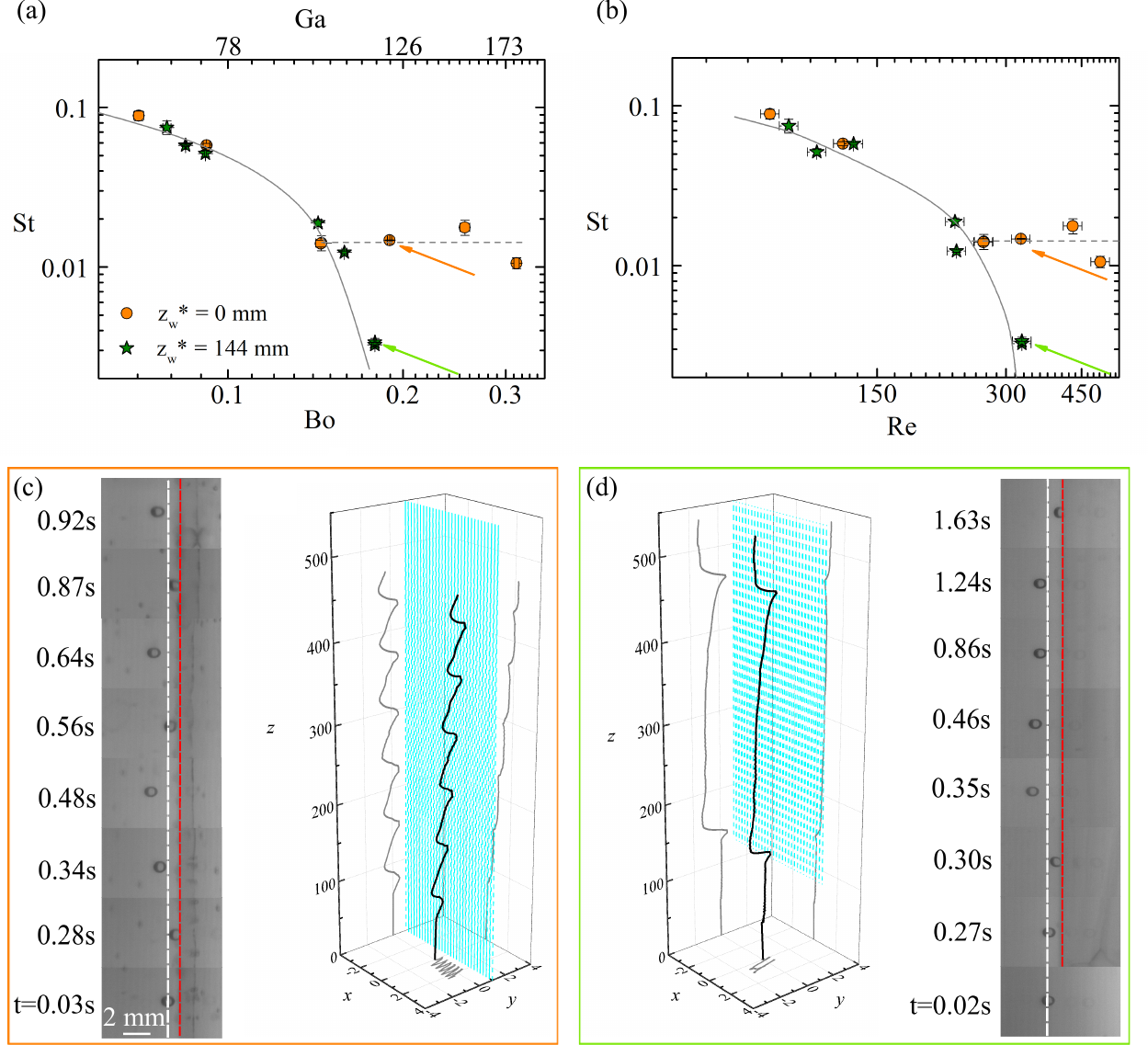}} 
\end{center}
\caption{Strouhal number $St$ versus Bond number $Bo$ and Galilei number $Ga$ for a wall present from the onset ($z_w^* = 0$ mm, orange circles) and located downstream ($z_w^* = 144$ mm, dark green stars) at $L=1$ in the PB regime. Solid and dashed lines indicate trend lines (a). Both datasets initially follow the trends $St\propto Bo^{-0.086}$ and $St \propto Re^{-0.055}$. However, the $z_w^* = 0$ mm case departs from this behaviour beyond $Bo \approx 0.15$ ($Re\approx 250$), reaching an approximately constant value around $St \approx 0.014$. (b) Strouhal number $St$ versus Reynolds number $Re$ for the same cases. (c) Time sequence illustrating the bubble dynamics for the $z_w^* = 0$ mm case, and corresponding dimensionless trajectory of the bubble, $Bo\approx0.19$, $Re\approx325$ and $St\approx0.015$. This case is associated with the reduction in Strouhal number observed in (a) and (b). (d) Dimensionless trajectory for the $z_w^* = 144$ mm case of the bubble, $Bo\approx0.18$, $Re\approx326$ and $St\approx0.0033$(d), and representative time sequence of the bubble evolution for the same case, corresponding to the plateau regime identified in (a). The white dashed line in the image sequences marks the $y = 0$ reference and spans the streamwise ($z$) direction, and the red dashed line indicates the vertical edge of the wall. The cases analyzed in panels (c) and (d) are highlighted by the orange and green arrows, respectively. Note that only one of the two overlapping data points indicated by the arrows is shown.} \label{fig:map_st}
\end{figure}

\subsection{Effect of the wall leading edge on the PB regime} \label{subsec:PB}
Once investigated the effect of the wall leading edge on the bubble's rising regime, we focus on the PB regime. In particular, we examine whether the bubble rebound frequency, $f^*$, is also influenced by the wall leading edge. The rebound frequencies were determined for each case by measuring the wavelength of the bubble trajectory and evaluating the time interval between successive rebounds, excluding the initial rebounds whenever possible. A rebound was identified from successive minima of the wall-normal centroid position. The period was computed as the mean time interval between consecutive minima after the initial transient. At least 2 cycles were used whenever available. The reported uncertainty combines the cycle-to-cycle variability and the standard deviation over 5 repetitions.

Figure~\ref{fig:map_st} presents the rebound frequency of bubbles undergoing periodic bouncing, expressed in terms of the Strouhal number, $St=f^*D^/v_t^*$, for the two wall configurations investigated. Both configurations follow a similar trend at low $Bo$, in particular, $St$ decreases with $Bo$ and $Re$. The increase in the time interval between successive collisions is consistent with the larger rebound amplitudes observed as bubble deformability increases~\citep{GRBV26}, indicating that the bubble spends progressively longer away from the wall before the next collision. Consequently, the rebound period $T^*=1/f^{*}$ becomes progressively larger relative to the characteristic convective time scale $D^*/v_t^*$, leading to a reduction in the Strouhal number. 

\begin{figure}[tbp]
\begin{center}
\resizebox{0.95\textwidth}{!}{\includegraphics{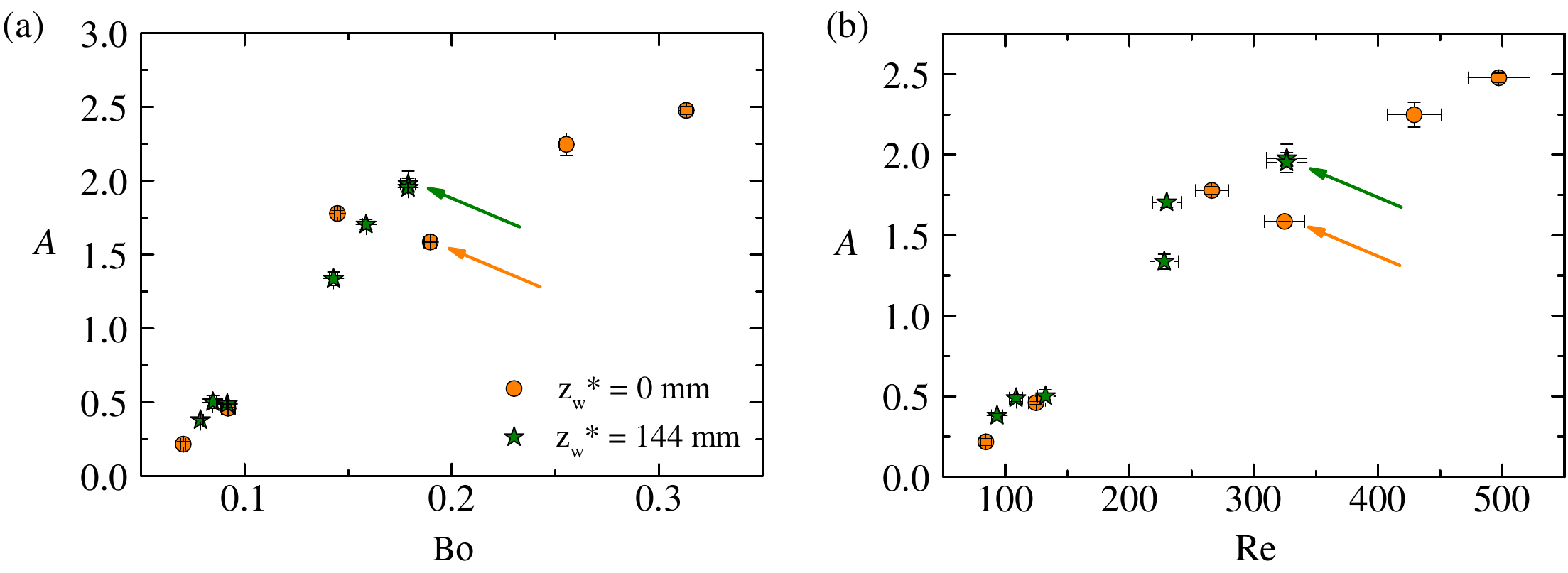}}
\end{center} 
\caption{Bubble rebound dimensionless amplitude $(A=A^*/D^*)$ as a function of Bond number, $Bo$ (a) and Reynolds number, $Re$ (b) in the PB regime. Orange circles correspond to the wall present from the onset ($z_w^*=0$ mm), whereas dark green stars correspond to the wall located downstream ($z_w^*=144$ mm), and the points of interest of the Figure \ref{fig:map_st} are marked with a orange arrow for $z_w^*=0$ mm,  $Bo\approx0.19$, $Re\approx325$ and $St\approx0.015$, and a green arrow for $z_w^*=144$ mm, $Bo\approx0.18$, $Re\approx326$ and $St\approx0.0033$.}
\label{fig:Ampl}
\end{figure}

\begin{figure}[tbp]
\begin{center}
\resizebox{0.99\textwidth}{!}{\includegraphics{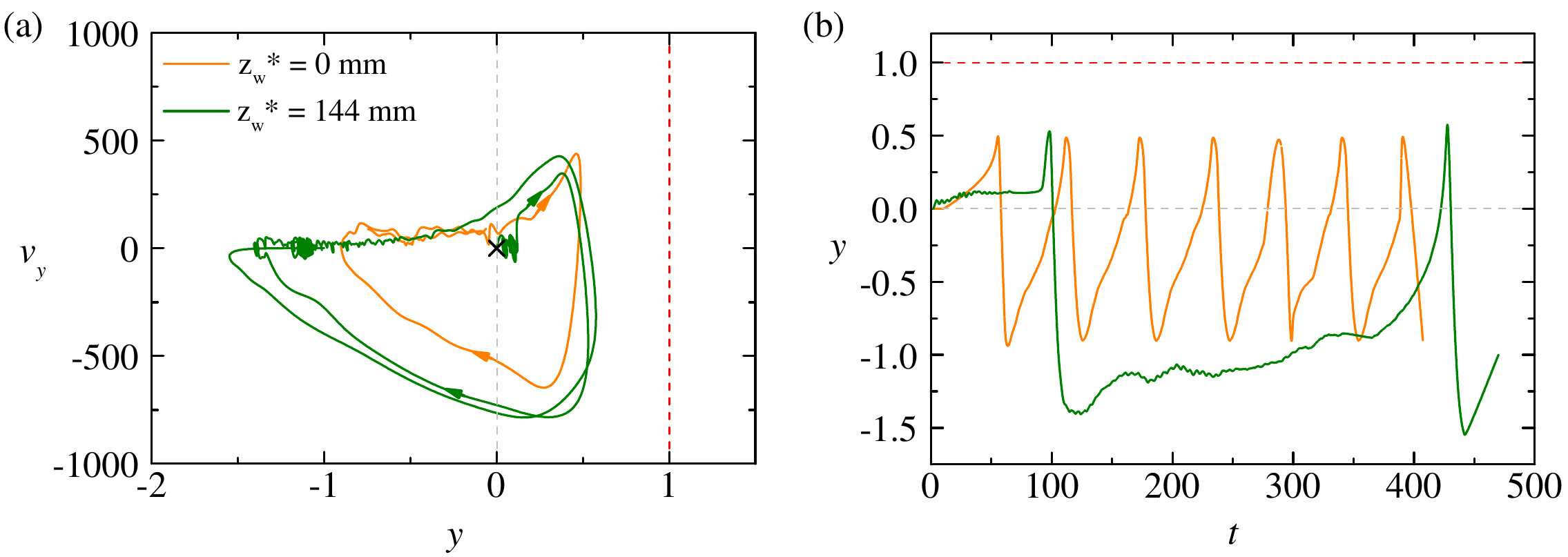}}
\end{center} 
\caption{(a) Dimensionless horizontal bubble velocity, $v_y=v_y^*/v_g$, where $v_g=\sqrt{g  D^*}$, as a function of the dimensionless transverse coordinate, $y=y^*/D^*$. The orange and dark green lines correspond to the bubbles shown in Figures \ref{fig:map_st} (c) and (d), with $z_w^*=0$ mm ($Bo\approx0.19$, $Re\approx325$ and $St\approx0.015$) and $z_w^*=144$ mm ($Bo\approx0.18$, $Re\approx326$ and $St\approx0.0033$), respectively. Only the last rebound loop is shown for the orange case. The beginning of the cycle is indicated by a black cross, and the arrows show the direction of the movement. The grey dashed line marks the position of the injector, $y = 0$, and the red dashed line indicates the vertical edge of the wall. (b) Dimensionless transverse coordinate, $y=y^*/D^*$, as a function of the dimensionless time $t=t^*v_t^*/D^*$. The complete trajectories are shown for both cases.} 
\label{fig:vxL}
\end{figure}

Nevertheless, a clear divergence between both configurations appears for ($Bo \gtrsim 0.15$, $Re \gtrsim 250$). In this range, the measured Strouhal number depends on the position of the wall leading edge: while it remains nearly constant for $z_w^*=0$, it decreases significantly for $z_w^*=144$ mm. For small Bond numbers, the bubble–wall interaction appears to be relatively insensitive to the conditions established before the first collision. Both wall configurations therefore converge towards essentially the same periodic dynamics. As the Bond number increases, the rebound becomes progressively more influenced by wake-induced inertial effects. For $Bo\gtrsim 0.15$ ($Re\gtrsim 250$), wake-induced inertial effects become sufficiently strong for differences in the pre-interaction wake state to persist over successive rebound cycles.

Figures \ref{fig:map_st}(c) and (d) show the paths of two bubbles with similar $Bo$ and $Re$ but very different Strouhal numbers depending on the vertical position of the wall. It can be noticed that the bubble rebound wavelength and amplitude are much lower for $z_w^*=0$ mm (Fig.~\ref{fig:map_st}c) than for $z_w^*=144$ mm (Fig.~\ref{fig:map_st}d), which implies a larger rebound frequency, and, therefore, higher $St$, for the $z_w^*=0$ configuration. These observations indicate that the periodic bouncing dynamics retains a memory of the conditions established before the first bubble--wall interaction and that the wall leading-edge position acts as an additional parameter influencing near-wall bubble dynamics.

The decrease in the Strouhal number observed for the downstream-wall configuration reflects a fundamental modification of the rebound cycle rather than an isolated change in the oscillation frequency. As shown below, the bubble reaches larger lateral distances from the wall, spends a considerably longer time returning towards it, and therefore completes each rebound cycle over a much longer time interval. The increase in rebound amplitude, the longer return stage and the reduction in Strouhal number should thus be regarded as different manifestations of the same underlying modification of the bubble dynamics induced by the wall leading-edge position.

Figure \ref{fig:Ampl} shows that the dimensionless rebound amplitude $A$ increases slightly with $Bo$ and $Re$. More importantly, for comparable values of both parameters and $Bo\gtrsim 0.15$ ($Re \gtrsim 250$), the downstream-wall configuration exhibits a larger rebound amplitude. This observation is fully consistent with the lower Strouhal numbers shown in Fig.~\ref{fig:map_st}(a)-(b), since a larger lateral excursion naturally requires a longer time before the bubble returns to the wall. The divergence between both configurations appears at approximately the same Bond number ($Bo\gtrsim0.15$) at which the PB-BTE transition also begins to differ, suggesting that both phenomena originate from the increasing influence of wake-induced inertial effects. 

The trajectories shown in Fig.~\ref{fig:vxL}(a) reveal why the oscillation period differs between both configurations. Although the maximum wall-normal velocities remain of the same order, the downstream-wall case (green line) undergoes a much more extended departure stage after collision. Consequently, the bubble travels farther from the wall and requires substantially longer to complete the return motion. This difference becomes particularly evident in Fig.~\ref{fig:vxL}(b), where the dimensionless rebound period is approximately 4.5 times larger than in the wall-from-injector configuration, in agreement with the ratio of the measured Strouhal numbers ($0.015/0.0033\approx4.5$). Therefore, the reduction in $St$ originates primarily from the longer duration of the rebound cycle rather than from significant changes in the characteristic transverse velocity.

The plateau value measured for the wall-from-injector configuration ($St\simeq0.014$) is remarkably close to the lowest frequencies reported in the DNS of \citet{SZM24}. In their simulations, the reduced frequency remains of order ($St\simeq0.11{-}0.12$) for weakly deformed bubbles, for instance at $Bo=4Bo_R\simeq0.08{-}0.20$. However, \citet{SZM24} also reported a pronounced reduction of the bouncing frequency in part of the periodic-bouncing region, with values as low as $St\simeq0.015$ for $Bo\simeq0.4$ for similar values of $Bo$ and $Ga$. This agreement suggests that the low-$St$ plateau observed here should not be interpreted as an anomalous frequency, but rather as a signature of a bouncing cycle with a long characteristic period. In the interpretation proposed by \citet{SZM24}, the oscillation frequency results from a balance between the effective stiffness of the transverse force near the equilibrium position and the virtual mass of the bubble–fluid system. As the bubble becomes more deformable, the equilibrium position shifts away from the wall and the lateral excursion increases, reducing the effective stiffness of the transverse restoring force and increasing the time required for the bubble to return to the wall. This mechanism is consistent with the decrease of $St$ observed in the present experiments up to $Bo\simeq0.15$. The subsequent plateau found for $z_w^*=0$ suggests that, once the wall-induced wake and vortex-shedding cycle are established from the beginning of the rise, the system selects an approximately constant minimum frequency. By contrast, when the wall is located downstream, the bubble first develops an almost axisymmetric wake before encountering the wall, and the subsequent reorganisation of the wake leads to a different bouncing dynamics, with lower frequencies or even transition to the escape regime. Although the comparison with the numerical data is necessarily qualitative since \citet{SZM24} varied $Bo$ and $Ga$ independently, whereas the present experiments are performed at fixed $Mo$, it supports the view that the rebound frequency is not fixed solely by $Bo$, $Ga$, $Re$ and $L$, but also by the initial wake configuration imposed by the wall leading edge position.

In addition, \citet{SZM25} reported Strouhal numbers in water of approximately ($St\approx0.05$) at ($Bo\approx0.6$), which are substantially larger than the plateau value observed here ($St\approx0.014$). However, these results correspond to a different dynamical regime, namely path oscillations associated with the near-wall zigzag instability, rather than the periodic bouncing regime investigated in the present work. Therefore, this discrepancy should not be interpreted as a disagreement between the two studies. Instead, it highlights that the characteristic frequency selected by periodic bouncing is considerably lower than that associated with large-amplitude path oscillations. This distinction suggests that the mechanisms controlling the oscillation period differ between the two regimes, despite both involving sustained lateral motions in the vicinity of a wall.

Finally, the present results may also help explain the discrepancy between the rebound frequency measured by \citet{TM03} and that predicted in the recent simulations of \citet{SZM24}. For $Mo=1.6\times10^{-9}$ and $Bo_R=0.073$, \citet{TM03} reported ($St\simeq0.06$), whereas the DNS of \citet{SZM24}, performed with a wall extending from the bubble injector, yielded a significantly larger value, ($St\simeq0.12$). Note that $Bo_R$ is the Bond number based on the bubble equivalent radius. Although several factors may contribute to this difference, including experimental uncertainties, slight differences in the initial bubble--wall separation, wall alignment, or contamination effects, the present measurements show that the wall leading-edge position alone can substantially modify the rebound frequency in the periodic-bouncing regime.
This suggests that the discrepancy between the experimental value of \citet{TM03} and the DNS prediction of \citet{SZM25} may not be solely due to numerical or experimental limitations, but may partly reflect a difference in the effective hydrodynamic configuration.

\section{Concluding Remarks}
\label{sec4}

The present work has experimentally investigated the influence of the wall leading-edge position on the interaction between a deformable bubble and a vertical wall. Two wall configurations were considered: a reference case in which the wall extends from the bubble injection point ($z^*_w=0$), and a second configuration in which the wall begins sufficiently far downstream for the bubble to attain its terminal rising state before entering the wall-bounded region ($z^*_w=144$ mm).

The experiments performed with the wall extending from the injection point show good qualitative agreement with the recent DNS of \citet{SZM25} regarding the sequence of interaction regimes and the transition from PB to BTE regime. This agreement provides experimental support for the numerical predictions obtained under equivalent wall configurations.

A direct comparison between the two wall configurations demonstrates that the wall leading-edge position has a pronounced influence on the transition between interaction regimes. When the wall begins downstream of the injector, the onset of the BTE regime is shifted towards smaller bubble sizes along the constant-Morton-number path, corresponding simultaneously to lower $Bo$, $Ga$ and $Re$. These observations support the hypothesis that differences in the wall configuration prior to the first bubble--wall interaction are responsible for a substantial part of the discrepancies previously reported between experimental observations and numerical simulations.

The wall leading-edge position also significantly modifies the dynamics within the periodic-bouncing regime. For $Bo\gtrsim 0.15$, when the bubble first rises freely before interacting with the wall, the rebound amplitude increases whereas the $St$ number decreases. Phase-space analysis reveals that this reduction in rebound frequency is not associated with substantially larger wall-normal velocities, but with a different rebound kinematics in which the bubble reaches larger wall-normal distances before returning towards the wall, leading to a significantly longer oscillation period.

Overall, the present results show that the hydrodynamic conditions established before the first bubble--wall interaction can influence the subsequent interaction dynamics. 

The present results demonstrate that the wall leading edge should be regarded as an additional control parameter in near-wall bubble dynamics. Although it does not modify the instantaneous bubble-wall separation, it changes the hydrodynamic history before the first interaction and thereby alters both the transition between interaction regimes and the characteristic rebound dynamics. These findings extend the current understanding of bubble--wall interactions and highlight the need to account for the complete interaction history when analysing or modelling bubbly flows involving finite walls or downstream obstacles.

\vspace{1cm}

This work was co-funded (85\%) by the European Union through the European Regional Development Fund (FEDER) and the Regional Government of Extremadura. Managing Authority: Ministry of Finance (Grant GR24077). Support from the Spanish Ministry of Science and Innovation (MCIN) through grant PID2022-140951OB-C22/ AEI/10.13039/501100011033/ FEDER, UE is gratefully acknowledged. R. Bola\~nos-Jim\'enez would like to acknowledge the project R1C\_2025\_011 funded by the Research and Knowledge Transfer Internal Plan of the University of Ja\'en. Finally, we thank Pengyu Shi for his helpful previous numerical simulations and discussion on the results.

\vspace{1cm}



\bibliographystyle{elsarticle-harv}
\bibliography{central,central2}

@ARTICLE{FMC07,
    Author         = {C. Ferrera and J. M. Montanero and M. G. Cabezas },
    Journal        = {Meas. Sci. Technol.},
    Pages          = {3713-3723},
    Title          = {An analysis of the sensitivity of pendant drops and liquid bridges to measure the interfacial tension},
    Volume         = {18},
    Year           = {2007}
}

@ARTICLE{LWZGZXLL22,
    Author         = {Y. Luo and Z. Wang and B. Zhang and K. Guo and L. Zheng and W. Xiang and H. Liu and C. Liu},
    Journal        = {Ind. Eng. Chem. Res. },
    Pages          = {9514-9527},
    Title          = {Experimental Study of the Effect of the Surfactant on the Single Bubble Rising in Stagnant Surfactant Solutions and a Mathematical Model for the Bubble Motion},
    Volume         = {61},
    Year           = {2022}
}

@ARTICLE{MM02,
    Author         = {G. Mougin and J. Magnaudet},
    Journal        = {Phys. Rev. Lett.},
    Pages          = {014502},
    Title          = {Path Instability of a Rising Bubble},
    Volume         = {88},
    Year           = {2002}
}

@ARTICLE{TSG15,
    Author         = {M. K. Tripathi and K. C. Sahu and R. Govindarajan},
    Journal        = {Nat Commun},
    Pages          = {6268},
    Title          = {Dynamics of an initially spherical bubble rising in quiescent liquid},
    Volume         = {6},
    Year           = {2015}
}

@ARTICLE{VBW02,
    Author         = {A.W.G {de Vries} and A. Biesheuvel and L. {van Wijngaarden}},
    Journal        = {Int. J. Multiphase Flow},
    Pages          = {1823-1835},
    Title          = {Notes on the path and wake of a gas bubble rising in pure water},
    Volume         = {28},
    Year           = {2002}
}

@ARTICLE{ZM08,
    Author         = {R. Zenit and J. Magnaudet},
    Journal        = {Phys. Fluids},
    Pages          = {061702},
    Title          = {Path instability of rising spheroidal air bubbles: A shape-controlled process},
    Volume         = {20},
    Year           = {2008}
}

@ARTICLE{CMMT16,
    Author         = {J. C. Cano-Lozano and C. Mart\'{\i}nez-Baz\'an and J. Magnaudet and J. Tchoufag},
    Journal        = {Phys. Rev. Fluids},
    Pages          = {053604},
    Title          = {Paths and wakes of deformable nearly spheroidal rising bubbles close to the transition to path instability},
    Volume         = {1},
    Year           = {2016}
}

@ARTICLE{BSFM24,
    Author         = {P. Bonneﬁs and J. Sierra-Ausin and D. Fabre and J. Magnaudet},
    Journal        = {J. Fluid Mech.},
    Pages          = {A19},
    Title          = {Path instability of deformable bubbles rising in Newtonian liquids: a linear study},
    Volume         = {980},
    Year           = {2024}
}

@ARTICLE{RVCMLH24,
    Author         = {A. Rubio and E.J. Vega and M.G. Cabezas and J. M. Montanero and J. M. López-Herrera and M. A. Herrada},
    Journal        = {Physics of Fluids},
    Pages          = {062112},
    Title          = {Bubble rising in the presence of a surfactant at very low concentrations},
    Volume         = {36},
    Year           = {2024}
}

@BOOK{CGW78,
    Address        = {USA},
    Author         = {R. Clift and J. R. Grace and M. E. Weber},
    Publisher      = {Academic Press},
    Title          = {Bubbles, Drops and Particles},
    Year           = {1978}
}

@ARTICLE{TM03,
    Author         = {F. Takemura and J. Magnaudet},
    Journal        = {J. Fluid Mech.},
    Pages          = {235-253},
    Title          = {The transverse force on clean and contaminated bubbles rising near a vertical wall at moderate Reynolds number},
    Volume         = {495},
    Year           = {2003}
}

@ARTICLE{SZM24,
    Author         = {P. Shi and J. Zhang and J. Magnaudet},
    Journal        = {J. Fluid Mech.},
    Pages          = {A8},
    Title          = {Lateral migration and bouncing of a deformable bubble rising near a vertical wall. Part 1. Moderately inertial regimes},
    Volume         = {998},
    Year           = {2024}
}

@ARTICLE{Shi24,
    Author         = {P. Shi},
    Journal        = {Phys. Rev. Fluids},
    Pages          = {023601},
    Title          = {Reversal of the transverse force on a spherical bubble rising close to a vertical wall at moderate-to-high Reynolds numbers},
    Volume         = {9},
    Year           = {2024}
}

@ARTICLE{TTMM02,
    Author         = {F. Takemura and S. Takagi and J. Magnaudet. Y. Matsumoto},
    Journal        = {J. Fluid Mech.},
    Pages          = {277-300},
    Title          = {Drag and lift forces on a bubble rising near a vertical wall in a viscous liquid},
    Volume         = {461},
    Year           = {2002}
}

@article{YPT03,
	author = {B. Yang and A. Prosperetti and S. Takagi},
	issn = {10706631},
	issue = {9},
	journal = {Phys. Fluids},
	pages = {2640-2648},
	title = {The transient rise of a bubble subject to shape or volume changes},
	volume = {15},
	year = {2003},
}

@article{M03c,
	author = {J. Magnaudet},
	issn = {00221120},
	issue = {485},
	journal = {J. Fluid Mech.},
	pages = {115-142},
	title = {Small inertial effects on a spherical bubble, drop or particle moving near a wall in a time-dependent linear flow},
	volume = {485},
	year = {2003},
}

@article{ST10,
	title={On the lateral migration of a slightly deformed bubble rising near a vertical plane wall},
	author={K. Sugiyama and F. Takemura},
	journal={J. Fluid Mech.},
	volume={662},
	pages={209-231},
	year={2010},
	publisher={Cambridge University Press}
}

@article{ZDCY20,
	author = {Y. Zhang and S. Dabiri and K. Chen and Y. You},
	issn = {0142-727X},
	journal = {Int. J. Heat Fluid Flow},
	month = {10},
	pages = {108649},
	publisher = {Elsevier},
	title = {An initially spherical bubble rising near a vertical wall},
	volume = {85},
	year = {2020},
}

@article{BM95,
	title={The structure of the axisymmetric high-{R}eynolds number flow around an ellipsoidal bubble of fixed shape},
	author={A. Blanco and J. Magnaudet},
	journal={Phys. Fluids},
	volume={7},
	number={6},
	pages={1265-1274},
	year={1995},
	publisher={American Institute of Physics}
}

@article{ST15,
	title={Direct numerical simulations of drag and lift forces acting on a spherical bubble near a plane wall},
	author={K. Sugioka and T. Tsukada},
	journal={Int. J. Multiph. Flow},
	volume={71},
	pages={32-37},
	year={2015},
	publisher={Elsevier}
}

@article{ECMBBJ25,
	title={Hydrodynamic forces on high Bond bubbles rising near a vertical wall at moderate Reynolds numbers: An experimental approach},
	author={C. Estepa-Cantero and C. Mart{\'\i}nez-Baz{\'a}n and R. Bola{\~n}os-Jim{\'e}nez},
	journal={Int. J. Multiph. Flow},
	volume={191},
	pages={105325},
	year={2025},
}

@ARTICLE{SZM25,
    Author         = {P. Shi and J. Zhang and J. Magnaudet},
    Journal        = {J. Fluid Mech.},
    Pages          = {A19},
    Title          = {Lateral migration and bouncing of a deformable bubble rising near a vertical wall. Part 2. Highly inertial regimes},
    Volume         = {1013},
    Year           = {2025}
}

@ARTICLE{GRBV26,
    Author         = {T. Gonz\'alez-Rubio and A. Rubio-Gonz\'alez and R. Bola{\~n}os-Jim\'enez and E. J. Vega},
    Journal        = {Int. J. Multiph. Flow},
    Pages          = {105723},
    Title          = {Effect of a downstream vertical wall on the path of a stable bubble rising at moderate-to-high Reynolds numbers: An experimental study},
    Volume         = {200},
    Year           = {2026}
}

\end{document}